\documentclass[aps,prl,twocolumn,showpacs,groupedaddress]{revtex4}
\usepackage{amsmath,amssymb,xspace,graphicx}

\newcommand\calp{{\cal P}}

\newcommand\calf{{\cal F}}

\newcommand\eff{\mbox{\rm eff}}
\newcommand\inter{\mbox{\rm int}}

\newcommand{\red}[1]{#1}

\begin{document}

\title{
Statistical Mechanics of Interacting Run-and-Tumble Bacteria}

\author{J. Tailleur and M. E. Cates}
\affiliation{SUPA, School of Physics, University of Edinburgh, JCMB
Kings Buildings, Edinburgh EH9 3JZ, United Kingdom}

\date{\today}

\begin{abstract}
We consider self-propelled particles undergoing run-and-tumble
dynamics (as exhibited by {\em E. coli}) in one dimension. Building on
previous analyses at drift-diffusion level for the one-particle
density, we add both interactions and noise, enabling discussion of
domain formation by `self-trapping', and other collective
phenomena. Mapping onto detailed-balance systems is possible in
certain cases.
\end{abstract}

\maketitle

Several species of bacteria, including {\em Escherichia coli}, perform
self-propulsion by a sequence of `runs' -- periods of almost
straight-line motion at near-constant speed ($v$) -- punctuated by
sudden and rapid randomizations in direction, or `tumbles', occurring
stochastically with rate $\alpha$.  It is no surprise that the
resulting class of random walk gives a diffusive relaxation of the
number density at large scales \cite{berg}. The resulting diffusion
constant $\sim v^2/\alpha$, is vastly larger than that of non-swimming
particles undergoing pure thermal motion at room
temperature. Therefore, apart from, {\em e.g.}, the upper limit it
imposes on the duration of a straight run (set by rotational
diffusion), true Brownian motion can usually be ignored.

Because bacterial diffusion is {\em not} thermal, the steady-state
probability density cannot be written as $p_{s} \propto
\exp[-H/k_BT]$, with $H$ a Hamiltonian, even for a single
diffuser. The physicist's intuition can easily be led astray: for
instance, Refs.\cite{purcell, schnitzer, othmer} address models (of
chemotaxis) comprising noninteracting particles in 1D, with no
external forces, but $v(x),\alpha(x)$ functions of position
$x$. Instead of a uniform density, as would arise with {\em any}
force-free detailed-balance dynamics, one finds $p_s(x)\propto 1/v(x)$
%. Thus steady state properties depend on $v$ and $\alpha$ separately, not just via $D(x)=v^2/\alpha$ 
\cite{purcell}. 

Here we extend previous analyses of run-and-tumble motion to the
many-particle level, addressing the roles of noise and
interactions. These determine, for instance, the dynamic correlator of
run-and-tumble bacteria, which is measurable by light scattering at
low density \cite{scattering} and at higher density, in principle, by
particle-tracking microscopy \cite{microscopy}. Additionally,
particles for which $v,\alpha$ depend on the local density (either via
thermodynamic interactions such as depletion \cite{depletion}, or
kinetic effects such as collision-induced tumbles) could show
collective phenomena such as domain-formation or flocking. Such
effects have previously been addressed within models where a
self-propelled particle responds vectorially to the velocity of its
neighbors, by direct sensing or passive hydrodynamics
\cite{toner,vicsek,goldstein}. Below, we shall find, for
run-and-tumble dynamics, similar effects in even simpler cases when
only the {\em speed} of a particle is density-dependent.

In making the transition from a single particle to many, most
bacterial modelling approximates the number density by a simple
replacement $\rho = Np$ \cite{schnitzer,othmer}. But even for
noninteracting particles, $\rho$ (unlike $p$) is a fluctuating
quantity, and a full statistical mechanics must compute noise terms
for $\rho$. As seen below, these are not ad-hoc, but follow from the
run-and-tumble dynamics directly.

To allow relatively rigorous progress
% and avoid pitfalls to intuition, 
we work in 1-D throughout. For $d>1$, although good descriptions exist
at the one-particle level \cite{schnitzer, othmer}, we leave many-body
effects to future work. To avoid unwieldy equations for the
probability flux $J$, we assume that, in units where run-and-tumble
parameters such as $v,\alpha$ are $O(1)$ quantities, spatial gradients
of these and of $J$ are $\ll 1$. (We show below that this was also
implicit in \cite{schnitzer}.) This restriction also judiciously
avoids complicated memory effects arising from the time-retarded
response function of bacteria \cite{degennes,berg}, that can lead to
much reduced universality \cite{kafri}.

\red{We start below with the microscopic dynamics of a single particle. We show how, on
large scales, its probability density is governed by a
drift-diffusion equation, whose Langevin counterpart we then extend to
describe the time evolution of the particle density in an assembly of
interacting run-and-tumble particles. We then illustrate some dramatic
consequences of these interactions and finally address the role of
external fields such as gravity.}

\red{Let us define $R(x,t),L(x,t)$ the probability of finding a single
run-and-tumble particle at $x,t$ in a right- or left-moving state
\cite{foot1}, respectively. With discrete run speeds $v_{R,L}$ and
$R\leftrightarrow L$, and interconversion rates $\alpha_{R,L}/2$ that
depend on $x$,
%but with no further averaging or approximation,% 
we have (with prime denoting $\partial_x$)}
\begin{eqnarray}
\dot R &=& -(v_RR)'-\alpha_RR/2 +\alpha_LL/2 \label{Rdot}
\\
\dot L &=& (v_LL)'+\alpha_RR/2 -\alpha_LL/2 \label{Ldot}
\\
\dot p &=& - J' \label{rhodot}
\end{eqnarray}
Eq.(\ref{rhodot}) sums (\ref{Rdot},\ref{Ldot}) to relate the
single-particle probability density $p \equiv R+L$ to its current $J
\equiv v_RR-v_LL$. Noise is not present in (\ref{Rdot},\ref{Ldot}),
which already represent an exact master equation for the dynamics of one
particle.
%(Noise would however be needed if we wished to reinterpret $R,L$ as local densities in a system of {\em many} noninteracting particles; see below.)

The diffusive limit is found by a second time differentiation,
elimination of $R,L$ and their derivatives, and setting $\ddot p \to
0$ \red{\cite{schnitzer,othmer}}.
The outcome can always be viewed as a time-local `constitutive'
relation $J = J[p(x)]$ for use with the continuity equation
(\ref{rhodot}). $J$, though linear in $p$, need not be local in space;
in fact setting $\ddot p = 0$ (which implies $\dot J = 0$) gives the
differential equation (with $2v \equiv v_L+v_R$ and $2\alpha \equiv
\alpha_L+\alpha_R$):
\begin{eqnarray}
(1&+&\xi_1)J+ \xi_2J'\label{JS1} = {\cal J}\\
\red{{\cal J}}&\red{\equiv}&\red{ - {\cal D}p' + {\cal V}p;}\qquad \label{JS2}
%\\
%\end{eqnarray}
%where 
%\begin{eqnarray}
\red{{\cal D}} \red{\equiv}\red{ v_Rv_L/\alpha} \label{defD}
\\
{\cal V} &\equiv& (\alpha_Lv_R-\alpha_Rv_L)/2\alpha - v(v_Rv_L/v)'/\alpha \label{defV}
%\end{eqnarray}
%\begin{eqnarray}
\\
\xi_1 &\equiv& [v_R(v_R/v)'-v_L(v_L/v)']/2\alpha\label{defxi1}\\
\xi_2 &\equiv& (v_R-v_L)/\alpha \label{defxi2}
\end{eqnarray} 
\if{
\begin{eqnarray}
 J&+& [v_R(v_RJ/v)'-v_L(v_LJ/v)']/2\alpha\label{JS1} = J_S\\
J_S&\equiv& - {\cal D}\rho' + {\cal V}\rho \label{JS2}
\\
%\end{eqnarray}
%where 
%\begin{eqnarray}
D_S &\equiv& v_Rv_L/\alpha \label{defD}\\
V_S &\equiv& (\alpha_Lv_R-\alpha_Rv_L)/2\alpha - v(v_Rv_L/v)'/\alpha \label{defV}
\end{eqnarray} 
}\fi An explicit nonlocal form $J[p]$ is readily found from
(\ref{JS1}). \red{By neglecting $J'$ in (\ref{JS1}), we
instead obtain the local form $J={\cal J}/(1+\xi_1)$ which is valid
as long as ${\cal V}/{\cal D}$ is itself small, a condition that must
anyway hold if our diffusion-drift description is to avoid large $p$
gradients in the flux-free steady state $p_s(x)$. The latter obeys
%\begin{equation}
$
{\cal V}/{\cal D} = (\ln p_s)' %\label{steadystate}
$,
%\end{equation} 
which follows exactly from (\ref{Rdot},\ref{Ldot}). Our local approximation reduces to that of Schnitzer~\cite{schnitzer} on further neglecting
$\xi_1$.  While the latter maintains $p_s$ exactly, by retaining $\xi_1$ we extend this exactness to {\em all} steady states
($J'=0$). 

We thus write as an optimal local approximation}
%
%\begin{eqnarray}
\begin{equation}
\red{\dot p = -J';\qquad J = -D p' + Vp} \label{bestflux}%  \label{bestone}
%\end{eqnarray}
\end{equation}
with %diffusivity and drift velocity 
$D\equiv {\cal D}/(1+\xi_1)$ and $V\equiv {\cal V}/(1+\xi_1)$. 

To analyse the more complex physics arising at many-body level, we
first exactly recast \red{(\ref{bestflux})} as an Ito-Langevin
process for a trajectory $x_i(t)$:
\begin{equation}
\dot x_i = A(x_i) + C(x_i)L_i(t) \label{ito} 
\end{equation}
with $L_i(t)$ unit-variance Gaussian white noise, $C^2=2D$, $A =
V+\partial D/\partial x_i$, and $C,A$ evaluated at the start of each
time increment \cite{oksendal}. The latter (Ito) prescription eases
the extension to many particles $x_i(t)$ whose parameters
$v_{R,L},\alpha_{R,L}$ depend on position, not only explicitly as in
(\ref{ito}), but also implicitly via the particle density. The latter
is formally constructed as $\rho(x) = \sum_i\rho_i(x) =
\sum_i\delta(x-x_i)$; a coarse-grained version (compatible with a
gradient expansion) is found by choosing a smoothed function for
$\rho_i(x)$. Within Ito's formulation \cite{oksendal}, no further
account need now be taken of the time-dependence of $\rho$: despite
interactions, the random displacements $C(x_i)L_i(t)\delta t$ depend
only on the {\em preceding} $\rho(t)$ and are statistically
independent of each other \cite{foot2}.

Accordingly, we can read Eq.(\ref{ito}) as the Ito-Langevin equations
for many interacting particles $i=1...N$, with $A = A(x,[\rho])$ and
$C = C(x,[\rho])$ \cite{footchop}. From these, the Ito-Langevin
equation for the collective density $\rho(x)$ is found by standard
procedures \cite{dean}, as follows. Ito's theorem \cite{oksendal}
states that for any function $f(x_i)$ of one variable
\begin{eqnarray}
\dot f(x_i)\! &=&\! (A+CL_i) \partial f/\partial x_i\!\! +\! (C^2/2) \partial^2f/\partial x_i^2\\
&=& \!\!\int \rho_i(x,t)[(A+C L_i)f' +Df'']dx \label{inter}
\end{eqnarray}
%where $L_i$ is not a function of $x$ but all other terms are. 
Integrating (\ref{inter}) by parts and using the identity
%\begin{equation}
$
\dot f(x_i)\equiv \int \dot\rho_i(x,t)f(x)dx \label{final}
$
%\end{equation}
gives 
\begin{equation}
\dot\rho_i = -(A\rho_i)' +(D\rho_i)'' - (L_iC\rho_i)' \label{single} 
\end{equation}
Summing on $i$ we obtain for the collective density \cite{footchop}
\begin{eqnarray}
\dot\rho &=& -(A\rho)' +(D\rho)''\!\! -\!\!\sum_i(CL_i\rho_i)' = - J_C'\label{multi}\\
J_C &=& \rho (V +  (\delta/\delta \rho)'D)- D\rho'+ (2D\rho)^{1/2}\Lambda\label{fluxform}
\end{eqnarray}
with $\langle\Lambda(x,t)\Lambda(x',t')\rangle =
\delta(t-t')\delta(x-x')$ \cite{MFT,chaikin,dean}. In
(\ref{fluxform}), $\rho$ is to be read as a coarse-grained, locally
smooth field.  Again by standard methods (but avoiding any appeal to
detailed balance) \cite{gardiner} the Fokker-Planck equation for the
many-body probability ${\cal P}[\rho]$ then follows as:
\begin{equation}
\dot{\cal P}\!\! =\!\!\! \int\!\!\! dx \frac{\delta}{\delta\rho(x)}\partial_x\!\!\left[\rho V\! -\!D\partial_x\rho\! -\! D\rho\!\left(\!\partial_x\frac{\delta}{\delta\rho(x)}\!\right)\right]\!{\cal P} \label{FPE}
\end{equation}

Allowing that $V,D$ are now functionals of $\rho$, we next seek
conditions under which we can map (\ref{FPE}), or equivalently
(\ref{ito}), onto a thermal system with detailed balance. (Clearly we
require no macroscopic flux, $\langle J_C\rangle = 0$.) In such a
system, forces derive from an excess free energy $\calf_{ex}[\rho]$;
diffusion and mobility matrices obey the Einstein relation $D_{ij} =
\mu_{ij}$; and steady state probability obeys $\calp_s[\rho]\propto
e^{-\calf[\rho]}$ with $\calf[\rho] = \calf_{ex} +
\int\rho(\ln\rho-1)dx$. (We set $k_BT = 1$ without loss of
generality.) Since in (\ref{ito}) the $L_i$ for different particles
are independent, $D_{ij}$ is diagonal \cite{foot2}.
%(This is also embodied in the form of (\ref{FPE}.) 
Hence the required condition is simply
\begin{equation}
V([\rho],x)/D([\rho],x) = -(\delta \calf_{ex}[\rho]/\delta \rho(x))' \label{integrable}
\end{equation} 
where the right hand side represents the force ({\em i.e.}, excess
chemical potential gradient) on a particle at $x$.

For 1D noninteracting particles, (\ref{integrable}) always holds, with
$\calf_{ex} = \int F(x)\rho(x)dx$ and $F(x) = \int
V(x')/D(x')dx'$. For instance, when $v_L\!=\!v_R=\!v(x)$ and
$\alpha_L\!=\!\alpha_R\!=\!\alpha(x)$, one has \cite{schnitzer,othmer} $F(x) =
\ln v(x)$ so that the mean steady-state density obeys $\rho_s(x) =
\rho_s(0)v(0)/v(x)$.
%(Clearly $F(x)$ is only {\em equivalent} to an excess free energy: $\ln v(x_i)$ is not really the free energy of an isolated run-and-tumble particle.)
%
For interacting particles, existence of $\calf_{ex}$ is not generic
even in 1D, as the configuration space is
$Nd$-dimensional. Nonetheless, some interesting cases do admit an
$\calf_{ex}$.  In particular, whenever $V = (s_1(\rho))'$ and $D =
s_2(\rho)$, with $s_{1,2}$ depending locally on $\rho$ only, then
$V/D$ satisfies (\ref{integrable}) with $\calf_{ex} = \int s_3(\rho)
dx$ and $s_3(\rho)$ obeying $s_2 d^2s_3/d\rho^2 = ds_1/d\rho$.

For example, consider a translationally invariant dynamics in which
$v_{R,L},\alpha_{R,L}$ at $x$ depend on $\rho(x), \rho'(x)...$, with
even parity: $\rho(x)\leftrightarrow\rho(-x)$ induces
$\alpha_L(x),v_L(x)\leftrightarrow \alpha_R(-x),v_R(-x)$.
%(Thus in a uniform state, $v_R = v_L$ and $\alpha_R=\alpha_L$.) 
Then
%\begin{equation}
$
(\delta \calf_{ex}[\rho]/\delta \rho(x))' = \Psi + (\ln(v_Rv_L/v))'
$
%\end{equation} 
where $\Psi \equiv (\alpha_Rv_L-\alpha_Lv_R)/2v_Rv_L$. On symmetry
grounds, $\Psi$ must vanish when all odd derivatives of $\rho$ do so;
thus to leading order in gradients we can write $\Psi =
(\psi(\rho))'$. Likewise $v_{R,L}-v \propto \pm\rho'$ so we can write
$(\ln(v_Rv_L/v))' =(\ln v(\rho))'$ to leading order. \red{Thus we recover
$\calf = \int f(\rho) dx=\int [\rho(\ln\rho-1)+f_{ex}(\rho)]dx$, with}
\begin{equation}
\red{ f_{ex}(\rho) =  \int_o^\rho[\psi(u) + \ln v(u)]du} \label{free} 
\end{equation}

To leading order in gradients, this system is then equivalent to a
system of Brownian particles at $k_BT = 1$, with mobility matrix
$D_{ij} = \delta_{ij}v^2(\rho(x_i))/\alpha(\rho(x_i))$, and a local
free energy density $f(\rho)$.
Thus, according to mean-field theory \cite{MFT}, whenever
\begin{equation}
\frac{d^2f}{d\rho^2} = \frac{1}{\rho}+\frac{d}{d\rho}[\psi(\rho) +\ln
v(\rho)] < 0
\end{equation}
the system is locally unstable toward spinodal decomposition into
domains of unequal density $\rho$. Also, it is globally unstable to
noise-induced (nucleated) phase separation whenever $\rho$ lies within
a common-tangent construction on $f(\rho)$ \cite{MFT,chaikin}.  In
particular, for $\psi = 0$ (left-right symmetry), any system with
$dv/d\rho<-v/\rho$ is liable, by the above reasoning, to undergo phase
separation. Since one result of finite tumble duration is effectively
to reduce $v$ \cite{foot1}, a strong enough tendency for the duration
of tumbles to increase with density may have similar effects.

In practice, of course, the very existence of a thermodynamic mapping
in this (strictly 1D) system ensures that the bulk phase separation
predicted by mean-field theory is replaced by Poisson-distributed
alternating domains of mean size $\propto e^\Delta$. Here $\Delta$ is
a domain-wall energy, fixed by gradient terms -- if these remain
compatible with the existence of ${\cal F}_{ex}$ (which
% in $f(\rho,\rho'...)$, although the continued existence of our thermal mapping at next order in gradients 
is not guaranteed).
To gain a first estimate of such gradient terms (retaining $\psi = 0$)
we choose a model where, in a uniform system, $v(\rho) =
v_oe^{-\lambda\rho}$ with $\lambda$ a constant. Now we argue that the
dependence $v$ on $\rho$ is somewhat nonlocal, sampling $\rho$ on
scales $\gamma$ of order the run distance, $v/\alpha$, which is $O(1)$
in our units. We therefore write for the nonuniform case $v =
v(\tilde\rho)$ where $\tilde\rho=\rho+\gamma^2\rho''$. (A linear term
is forbidden by symmetry.)  Thus $\delta\calf_{ex}/\delta\rho(x) = \ln
v = \ln v_o-\lambda(\rho+\gamma^2\rho'')$ and
\begin{equation}
f_{ex}(\rho,\rho') = \rho\ln v_o+\lambda [-\rho^2 + \gamma^2
\rho'^2]/2 \label{landau}
\end{equation}

The resulting $f(\rho)$ is very familiar \cite{chaikin}, and locally
unstable for all $\rho \ge 1/\lambda$.
A stable dense phase is however regained if $v(\rho)$ saturates at
large $\rho$. Suppose, {\em e.g.}, $v = v_o\exp[-\lambda \varphi
\arctan(\rho/\varphi)]$, which falls as $v_oe^{-\lambda\rho}$ before
approaching $v_{sat}= v_o \exp[-\lambda\varphi\pi/2]$ at $\rho \gg
\varphi$. So long as $\varphi> 2/\lambda$, a window of phase
separation is maintained in mean field. Notably, some of the gradient
terms arising from (the above form of) nonlocality now violate our
thermodynamic mapping. However, if one ignores such violations,
$\Delta$ is found to be large, and domain formation accordingly
pronounced, whenever $\varphi \gg 1$.

\begin{figure}
\centerline {\includegraphics{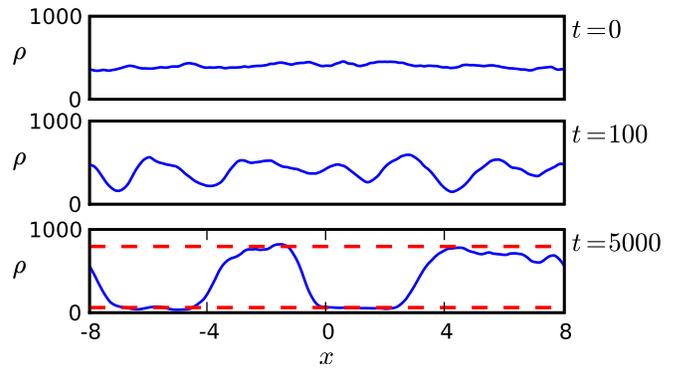}}
\caption{\red{Spinodal-like behavior of 6400 interacting particles
within the region where $d^2f/d\rho^2 < 0$. In the numerics,
$\tilde\rho$ is defined by convolution of $\rho$ with a smooth
function of finite range $\pm 1/2$; we set $v_0 = 2.5, \lambda = 0.01,
\varphi = 250$. The box length is $16$, with periodic boundary
conditions, and the mean density is $400$. Dashed lines show the
common-tangent densities ($\rho = 60; 800$).  }}
\label{fig}
\end{figure}

Although rigor is now exhausted, the physics seems reasonable: we know
\cite{purcell,schnitzer,othmer} that for an imposed $v=v(x)$ particles
accumulate in regions of low $v$. Thus with $dv/d\rho$ sufficiently
negative, `self-trapping' of high-density, slow-moving domains can be
expected.
\red{To investigate whether this scenario arises, we have simulated
Eqs.(\ref{ito}) for the above conditions and indeed observed the
predicted spinodal dynamics (Fig.\ref{fig}).} Moreover, if we create a
fully phase-separated initial state, this shows prolonged stability
when the mean density lies between the predicted binodals; outside
these, it collapses to uniformity. Thus the self-trapping scenario
appears valid despite violations of our thermodynamic mapping at
gradient level. \red{Eqs.(\ref{multi},\ref{fluxform})
generalize obviously to $d>1$, where they would lead to genuine phase
separation under similar conditions. However, an adequate local approximation relating $A,C$ to run-and-tumble parameters remains to be established, and the range of validity of the diffusive limit is unknown. In $d>1$ additional physics also enters, such as hydrodynamic interactions which are only partially accounted for
by our use of a density-dependent velocity field~\cite{goldstein}.}

%Of course it would be most interesting to know whether the same physics could, in $d>1$, lead to genuine phase separation.
%Of course, even in 1D this simplest of models represents only one corner of a parameter space likely to be dominated by regions admitting no mapping onto any thermal system. Nonetheless, finding such mappings, where they exist, offers a first route into studying the dynamics.

Finally, let us consider a translationally invariant system with no
density dependence of $\alpha$ nor of $v$, but where $v_{L,R}$ are
biased by some colloidal interaction $H_{\inter}$. That is,
$v_{R,L}(x) = v \pm v_T$, with $v_T = -\mu_T(\delta
H_{\inter}/\delta\rho(x))'$ and $\mu_T$ the mobility (inverse
friction) \cite{depletion}. To first order in small $v_T$ we then have
$V = v_T$ and $D = v^2/\alpha$. The latter swamps any small thermal
contribution ($D_T= k_BT\mu_T$), so that (\ref{ito},\ref{FPE}) are, to
this order, equivalent to a thermodynamic system with $H_{\rm int}$,
but at enhanced temperature $k_BT_{\eff} \simeq D/\mu_T$.  Correlators
such as ${\mathcal S}(q,t'-t)\equiv \langle \rho_q(t)\rho_{-q}(t')\rangle$
follow, although in many cases $T_{\rm eff}$ may be so large that
these approach the noninteracting limit, ${\mathcal S}(q,t) =
N\exp[-Dq^2t]$ \cite{foot}.  A similar expansion
shows $T_{\eff}$ also to control sedimentation equilibrium under weak
enough gravity ($H = H_{\inter} + mg\int \rho(x) x dx$).

For the chosen $\rho$-independent $v$ and $\alpha$, this effective
temperature picture is intuitively clear and appealing.
Nonperturbatively, however, ${\cal V} = v_T +(v_T^2)'/\alpha$ and
${\cal D} = (v^2-v_T^2)/\alpha$. Since $v_T = v_T[\rho]$,
Eq.(\ref{integrable}) no longer holds: the effective temperature
concept breaks down as soon as the colloidal or gravitational
interactions are non-infinitesimal. (This is true even within the
gradient expansion, which itself fails at $v_T/v \simeq 1$.)
Perhaps instructive is the exactly solved case of {\em noninteracting}
sedimentation, where $v_T = -\mu_T m g$. This gives in steady state a
Boltzmann-like exponential density, $\rho_s(x) =\rho(0) e^{- \kappa
x}$, whose spatial decay rate $\kappa = -v_T \alpha/(v^2-v_T^2)$ is,
however, nonlinear in $g$ \cite{trap}.
On increasing $g$, the density profile collapses to zero height, not
when $g\to\infty$, but when $v_T(g)\to v$. Although cut off by other
physics (such as true Brownian motion) this finite-$g$ singularity
occurs because for $|v_T|>v$, $L$ and $R$ particles move in the same
direction: a steady state of zero flux is clearly then
impossible. Strong colloidal forces may likewise create absorbing
states whose local `escape velocity' exceeds $v$.
%More generally this holds whenever $v$, even if variable, is bounded above (which in thermal equilibrium it is not).
This conclusion would alter if $v$ represented not a fixed value, but
the mean of a distribution extending (as in a truly thermal system) to
unlimited speeds. Existence of a strict upper limit to the speed of
self-propelled particles could thus be an important factor in their
physics.

In summary, by considering the passage from local run-and-tumble to
drift-diffusion dynamics at large scales, we have elucidated the roles
of both noise and interactions.
%This was done within a 1D gradient expansion. 
When the mean run speed $v$ is a sufficiently decreasing function of
local density (or tumble-time $\tau$ sufficiently increasing
\cite{foot1}), purely kinetic interactions could cause `self-trapping'
of domains in 1D, suggestive of bulk phase separation in higher
dimensions. For particles interacting not by kinetic but by
conventional thermodynamic forces (creating local drift velocities
superposed on a density-independent run-and-tumble dynamics) such a
mapping gives, perturbatively, an effective temperature set by the
ratio of the run-and-tumble diffusivity to the thermodynamic
mobility. However this mapping breaks down as soon as the drift
velocities $v_T$ arising from the interactions become a significant
fraction of the run-speed, $v$. For $|v_T|>v$, and in the absence of
true Brownian motion, absorbing states are possible.

Because of the progress it allows, we have focussed above on those
exceptional cases that, despite nonequilibrium interactions and noise,
admit a thermodynamic mapping. Accordingly, large areas of parameter
space remain unexplored; these could harbor many further interesting
forms of collective nonequilibrium behavior.

We thank R. Blythe, M. Evans, W. Poon and S. Ramaswamy (discussions)
and EPSRC EP/E030173 (funding). MEC holds a Royal Society Research
Professorship.


\begin{thebibliography}{99}
\bibitem{foot1} We take tumbles as instantaneous, but a finite mean duration $\tau$ is similar to setting $(v,\alpha)\to (v,\alpha)/(1+\alpha\tau)$. Indeed, for Poissonian  tumble durations and $v_R,\alpha_{R} =v_L,\alpha_L$, the stated diffusive limit is thereby recovered, on neglect of all second-order time derivatives.
\bibitem{foot2} Velocity-dependent forces, such as hydrodynamic interactions, would change this; see also \cite{goldstein}.
\bibitem{footchop} 
We have $A(x_i,[\rho]) = V(x_i,[\rho]) + \partial D(x_i,[\rho])/\partial x_i$, whose second term is found by the chain rule as $[D'(x) + (\delta/\delta \rho(x))'D(x)]_{x=x_i}$, and
leads directly to the $(\delta/\delta\rho)'D$ term
in (\ref{fluxform}). This vanishes whenever $D(x_i,[\rho])$ can be written $D = \tilde D(x_i,[\rho-\rho_i(x)])$, which excludes self-density interaction.
\bibitem{MFT} Omitting  the noise term from (\ref{fluxform}) as in 
\cite{schnitzer,othmer} amounts to a dynamic MFT (mean field theory), 
$J_C[\rho] = J_C(x,[\langle\rho\rangle]$, which loses the statistical mechanics of fluctuations and thus of activated phenomena such as nucleation. Static MFT instead finds the global extremum of ${\cal F}[\rho]$, allowing access to nucleated phase separations \cite{chaikin}. The thermodynamic mappings reported below include all noise terms and are therefore not contingent on either MFT. In some further cases, not explored here, such mappings can be made within dynamic MFT but are then violated by fluctuations.
\bibitem{dean} D. S. Dean, J. Phys. A Math. Gen., {\bf 29}, L613-L617 (1996); we extend this analysis here to the case where $A,C$ are functionals of $\rho$. See also P. H. Chavanis [arXiv:0803.0263]
\bibitem{foot} These results hold only at diffusive scales, $qv/\alpha\ll 1$, whereas the opposite limit is more easily reached in scattering experiments \cite{scattering}.
\bibitem{trap} Noninteracting particles in a quadratic potential, $v_{\tau} = - kx$, obey $\rho_s(x) = \rho_0|1-k^2x^2/v^2|^{\alpha/2k-1}$ which is even further from Boltzmann form.
\end{thebibliography}
\end{document}